\documentstyle[12pt,amssymb,epsfig]{article}
\begin{document}
\setlength{\parskip}{0.45cm}
\setlength{\baselineskip}{0.75cm}
\renewcommand{\thefootnote}{\fnsymbol{footnote}}
\begin{titlepage}
\begin{flushright}
DO-TH 97/23 \\ DTP/97/96  \\ November 1997
\end{flushright}
\vspace{0.6cm}
\begin{center}
\Large
\hbox to\textwidth{\hss
{\bf Mesonic Parton Densities Derived From }\hss}

\vspace{0.1cm}
\hbox to\textwidth{\hss
{\bf Constituent Quark Model Constraints} \hss}

\vspace{1.2cm}
\large
M.\ Gl\"{u}ck and E.\  Reya\\
\vspace{0.5cm}
\normalsize
Institut f\"{u}r Physik, Universit\"{a}t Dortmund, \\
\vspace{0.1cm}
D-44221 Dortmund, Germany \\
\vspace{1.2cm}
\large
M.\ Stratmann \\
\vspace{0.5cm}
\normalsize
Department of Physics, University of Durham, \\
\vspace{0.1cm}
Durham DH1 3LE, England \\
\vspace{1.6cm}
{\bf Abstract}
\vspace{-0.3cm}
\end{center}
Using constituent quark model constraints we calculate the
gluon and sea content of pions solely in terms of their 
valence density and the known sea and gluon densities of the
nucleon. The resulting small-$x$ predictions for $g^{\pi}(x,Q^2)$
and $\bar{q}^{\pi}(x,Q^2)$ are unique and parameter free, being entirely
due to QCD dynamics. Similar ideas are applied for calculating the gluon
and sea content of kaons which, for our suggested choice of the kaon's
valence densities, turn out to be identical to the ones of the pion.
\end{titlepage}
%
%MAIN PART
%
%INTRODUCTION
%
\section{Introduction}
\noindent
The parton content of the mesons, $\pi$, $K$, $\rho$, $\ldots$, is
not well known due to the scarce experimental 
information solely from Drell-Yan dilepton production processes
as compared to
the rich and accurate data which exist for the nucleon
from various different reactions.
One can try to improve the situation by relating the (rather) well known
nucleonic parton distributions to the poorly known mesonic ones
utilizing a plausible constituent quark description of the hadrons
in which the partons are considered as {\em{universal}} parts of the
constituent quarks [1-3]. This model was applied recently \cite{ref4}
to predict the pion structure from the known nucleon structure 
functions utilizing the constituent wave functions in \cite{ref1}.

As noted in \cite{ref4}, the choice of the constituent wave functions
introduces some ambiguity in the prediction of the pion structure
functions from those of the proton. We shall therefore employ a slightly 
different approach which eliminates the dependence on the constituent
wave functions without sacrificing completely the predictive power of the
model. In section 2 we apply our approach to the pion
in leading order (LO) and next-to-leading order (NLO) of
QCD, while section 3 is devoted to the $K$ meson.

%%%%%%%%%%%%%%%%%%%%%%%%%
% SECTION 2
%%%%%%%%%%%%%%%%%%%%%%%%%
\section{The Pion Structure}
Following refs.\ \cite{ref1,ref4}, the constituent building blocks
of the nucleon and the pion will be denoted by $U$ and $D$, i.e.\
$p=UUD$, $\pi^+=U\bar{D}$, etc., and their distributions within the
proton [pion] will be denoted by $U^p(x)$ [$U^{\pi}(x)$], etc., which
are scale ($Q^2$) independent. Their {\em{universal}} (i.e., hadron
independent) partonic content will be denoted by $v_c(x,Q^2)$,
$g_c(x,Q^2)$, and $\bar{q}_c(x,Q^2)$ representing the valence, gluon,
and sea components of the constituent $U^h$, $D^h$ distributions,
respectively.
The usual parton content of the proton is then given by
\setcounter{equation}{0}
\renewcommand{\theequation}{1\alph{equation}}
\begin{eqnarray}
u_v^p &=& U^p \otimes v_c \\
d_v^p &=& D^p \otimes v_c \\
\bar{q}^p &=& \left(U^p+D^p\right) \otimes \bar{q}_c \\
g^p &=& \left(U^p+D^p\right) \otimes g_c
\end{eqnarray}
where $u_v^p\equiv u^p-\bar{u}^p$ and $d_v^p\equiv d^p -
\bar{d}^p$ are the valence quark densities,
$\bar{q}^p=\left(\bar{u}^p+\bar{d}^p\right)/2$, and $\otimes$ 
denotes the usual convolution which becomes a simple
product for the corresponding Mellin $n$-moments, henceforth
utilized in our discussion and for our explicit calculations.
We rewrite these equations in Mellin $n$-moment space as follows:
\setcounter{equation}{0}
\renewcommand{\theequation}{2\alph{equation}}
\begin{eqnarray}
v^p &\equiv& u_v^p + d_v^p = \left(U^p+D^p\right) v_c \\
\bar{q}^p &=& \left(U^p+D^p\right) \bar{q}_c \\
g^p &=& \left(U^p+D^p\right) g_c
\end{eqnarray}
with $v^p\!=\!v^p(n,Q^2)\!\equiv\!\int_0^1 x^{n-1} \left[ u_v^p(x,Q^2)\!+\!
d_v^p(x,Q^2)\right] dx$,
$v_c\!=\!v_c(n,Q^2)\!\equiv\!\int_0^1 x^{n-1} v_c(x,Q^2) dx$, etc., and where
we omit the obvious $n$ and $Q^2$ dependence in $v^p(n,Q^2)$,
$v_c(n,Q^2)$, etc., whenever possible. Similarly we obtain for 
the pion\footnote[2]{It should be recalled that 
$u_v^{\pi^{+}}=\bar{d}_v^{\pi^{+}}=\bar{u}_v^{\pi^{-}}=
d_v^{\pi^{-}}$, $\bar{u}^{\pi^{+}}=d^{\pi^{+}}=u^{\pi^{-}}=
\bar{d}^{\pi^{-}}$ and $q^{\pi^{0}}=(q^{\pi^{+}}+
q^{\pi^{-}})/2$. Similarly,
$u_v^{K^{+}}=\bar{u}_v^{K^{-}}$, $\bar{s}_v^{K^{+}}=
s_v^{K^{-}}$ and $\bar{u}^{K^{+}}=u^{K^{-}}=d^{K^{\pm}}=
\bar{d}^{K^{\pm}}$.} 
\setcounter{equation}{0}
\renewcommand{\theequation}{3\alph{equation}}
\begin{eqnarray}
v^{\pi} &\equiv & u_v^{\pi^{+}}+\bar{d}_v^{\pi^{+}} = 
\left(U^{\pi^{+}}+\bar{D}^{\pi^{+}}\right) v_c \\
\bar{q}^{\pi} &=& \left(U^{\pi^{+}}+
\bar{D}^{\pi^{+}}\right) \bar{q}_c \\
g^{\pi} &=& \left(U^{\pi^{+}}+\bar{D}^{\pi^{+}}\right) g_c
\end{eqnarray}
with $\bar{q}^{\pi}=\left(\bar{u}^{\pi^{+}}+d^{\pi^{+}}\right)/2$
and $\bar{u}^{\pi^{+}}=d^{\pi^{+}}$ due to the common neglect
of ${\mathrm{SU(2)}}_{flavor}$ breaking effects in the $\pi$.
The above equations are {\em{conceived to apply at the low resolution 
scale}} $Q^2=\mu^2$ ($\mu_{LO}^2=0.23\,\mathrm{GeV}^2$,
$\mu_{NLO}^2=0.34\,\mathrm{GeV}^2$) of \cite{ref5} where the strange 
quark distribution is considered to be negligible, i.e., 
\setcounter{equation}{3}
\renewcommand{\theequation}{\arabic{equation}}
\begin{equation}
s^p(x,\mu^2) = \bar{s}^p(x,\mu^2) =0\;\;\;.
\end{equation}
We shall adopt the same approximation also for the pion, i.e.,
\begin{equation}
s^{\pi}(x,\mu^2) = \bar{s}^{\pi}(x,\mu^2) =0\;\;\;.
\end{equation}
Note that in contrast to our previous analysis \cite{ref6} of the
pion structure, we now start with a {\em{non}}-vanishing sea
$\bar{q}^{\pi}$ at $Q^2=\mu^2$ as follows from eqs.\ (2b) and (3b).

It is easily seen that eqs.\ (2) and (3) yield the following
{\em{wave function independent}} relations
\begin{equation}
\frac{v^{\pi}}{v^p} = \frac{\bar{q}^{\pi}}{\bar{q}^p} =
\frac{g^{\pi}}{g^p}
\end{equation}
which, together with eq.\ (5), fix the pion structure in terms of the
proton structure as soon as $v^{\pi}$ is reasonably well determined:
\begin{equation}
g^{\pi}=\frac{v^{\pi}}{v^p}\, g^p\;\;,\;\;
\bar{q}^{\pi}=\frac{v^{\pi}}{v^p}\, \bar{q}^p\;\;.
\end{equation}
These $n$-moment relations are our basic predictions for the gluon
and sea densities of the pion at the input scale $Q^2=\mu^2$. The
required LO and NLO input densities of the proton are taken from
\cite{ref5}, with $\bar{q}^p$ referring to the average of the
$\bar{u}$ and $\bar{d}$ sea densities of \cite{ref5}, i.e.,
$\bar{q}^p=\left(\bar{u}^p+\bar{d}^p\right)/2$.
Furthermore the sum rules
\begin{eqnarray}
\int_0^1 v^{\pi}(x,Q^2)\, dx &=& 2\\
\int_0^1 x v^{\pi}(x,Q^2)\, dx &=& \int_0^1 x v^p(x,Q^2)\, dx
\end{eqnarray}
impose strong constraints on $v^{\pi}(x,\mu^2)$ which are very
useful for its almost unambiguous determination from 
experimental Drell-Yan data in $\pi N$ collisions. Independent
analyses of the valence structure of protons and pions 
within the framework of the radiative parton model \cite{ref5,ref6} 
suggest that the valence quarks in the proton and the pion
carry similar total fractional momentum, as implied by eq.\ (9).
In practice we can therefore utilize the $v^{\pi}(x,\mu^2)$ of \cite{ref6}
slightly modified so as to comply with the new constraint
in eq.\ (9). This yields 
\begin{eqnarray}
v_{LO}^{\pi}(x,\mu_{LO}^2) &=& 0.942\, x^{-0.501} \left(
1+0.632 \sqrt{x}\right) \left(1-x\right)^{0.367}\\
v_{NLO}^{\pi}(x,\mu_{NLO}^2) &=& 1.052\, x^{-0.495} \left(
1+0.357 \sqrt{x} \right) \left(1-x\right)^{0.365}\;\;.
\end{eqnarray}
The total momentum fractions carried by these LO and NLO input
valence densities are given by
\begin{equation}
\int_0^1 x v_{LO}^{\pi}(x,\mu_{LO}^2) \, dx = 0.603\;\;,\;\;
\int_0^1 x v_{NLO}^{\pi}(x,\mu_{NLO}^2)\, dx = 0.582
\end{equation}
which coincide, as they should, with the ones 
of the proton \cite{ref5}, cf.\ eq.\ (9).

Having completely fixed the input for $g^{\pi}$ and 
$\bar{u}^{\pi^{+}}$ in eq.\ (7), we perform the LO and NLO 
evolutions of $g^{\pi}(n,Q^2)$ and $\bar{u}^{\pi^{+}}(n,Q^2)$ to
$Q^2>\mu^2$ in Mellin $n$-moment space, followed by a straightforward
numerical Mellin-inversion \cite{ref7} to Bjorken-$x$ space for
obtaining $g^{\pi}(x,Q^2)$ and $\bar{u}^{\pi^{+}}(x,Q^2)$. The same
is done for $s^{\pi}$, starting from the vanishing input in (5). It should
be noted that the evolutions are always performed in the fixed
(light) $f=3$ flavor factorization scheme \cite{ref5,ref8}, i.e.,
we refrain from generating radiatively massless 'heavy' quark
densities $h(x,Q^2)$ where $h=c,\,b,$ etc. Hence heavy quark
contributions have to be calculated in fixed-order perturbation
theory via, e.g., $g^{\pi} g^{p}\rightarrow h\bar{h}$,
$\bar{u}^{\pi} u^p \rightarrow h\bar{h}$, etc.

In fig.\ 1 we compare our present LO and NLO input parton distributions at
$Q^2=\mu^2$ with those of \cite{ref6}, while fig.\ 2 shows our 
resulting predictions for various larger fixed values of $Q^2$ as 
compared again to our \cite{ref6} former results denoted by
${\mathrm{GRV}}_{\pi}$. In contrast to our former \cite{ref6} SU(3)$_{flavor}$
symmetric sea $\bar{q}^{\pi}$, the present one is merely 
SU(2)$_{flavor}$ symmetric and $\bar{q}^{\pi}$ refers now to the
quantity in eq.\ (3b) while $s^{\pi}=\bar{s}^{\pi}$ is not shown in the
figure since it practically coincides with our previous 
GRV$_{\pi}$ \cite{ref6} $\bar{q}^{\pi}$ for the following
reasons:
Our unique parameter free small-$x$
$(x\lesssim 10^{-2})$ predictions for $xg^{\pi}$ and 
$x\bar{q}^{\pi}$ at $Q^2>\mu^2$ in fig.\ 2 are entirely due to 
QCD dynamics since they are radiatively generated from the 
valence-like input densities at $Q^2=\mu^2$
which vanish as $x\rightarrow 0$. Thus the results for $g^{\pi}$ 
and $\bar{q}^{\pi}$ almost coincide
with the ones of \cite{ref6} except for $x\bar{q}^{\pi}$ at
$x\gtrsim 10^{-2}$ which has been generated from a vanishing input
\cite{ref6}, in contrast to the present analysis.
Furthermore our predictions for $s^{\pi}$, resulting also from the
vanishing input in eq.\ (5), almost coincide with our previous
GRV$_{\pi}$ $\bar{q}^{\pi}$ shown in fig.\ 2.

In fig.\ 3 we present a more detailed comparison of our present 
(solid lines) and previous NLO \cite{ref6} results at a specific value
of $Q^2=20\,{\rm{GeV}}^2$ where we also show the corresponding 
distributions of \cite{ref9}. Finally our predictions are confronted 
in fig.\ 4 with a representative sample of the
experimental data \cite{ref10,ref9} from
the Drell-Yan process ($\pi^-W$ reactions) as also done in
\cite{ref9,ref4}. It should be noted that our NLO 
$K$-factors, i.e.\ $K'$, are similar to the ones obtained
in \cite{ref4,ref9}.
The relevant NLO differential Drell-Yan cross section 
$d^2\sigma/d\sqrt{\tau}dx_F$ has been presented in the 
Appendix of \cite{ref9} except for eq.\ (A8) which has to be
modified \cite{newref11,newref12} in order to conform with the
usual $\overline{\mathrm{MS}}$ convention for the number of gluon
polarization states $2(1-\epsilon)$ in $4-2\epsilon$ dimensions.

%%%%%%%%%%%%%%%%%%%%%%%%
% SECTION 3
%%%%%%%%%%%%%%%%%%%%%%%%
\section{The Kaon Structure}
Here eqs.\ (3a)-(3c) are obviously replaced by\footnotemark[2]
\setcounter{equation}{0}
\renewcommand{\theequation}{13\alph{equation}}
\begin{eqnarray}
v^K &\equiv & u_v^{K^{+}}+\bar{s}_v^{K^{+}} = \left(
U^{K^{+}}+\bar{S}^{K^{+}}\right) v_c \\
\bar{q}^K &=& \left(U^{K^{+}}+\bar{S}^{K^{+}}\right) 
\bar{q}_c\\
g^{K}&=& \left(U^{K^{+}}+\bar{S}^{K^{+}}\right) g_c
\end{eqnarray}
where as in the case of the pion, $\bar{q}^K \equiv  \bar{u}^{K^{+}}=
d^{K^{+}}=\bar{d}^{K^{+}}$, and for the strange sea input we take again
\setcounter{equation}{13}
\renewcommand{\theequation}{\arabic{equation}}
\begin{equation}
s^{K^{+}} = \bar{s}^{K^{-}} =0\;\;\;.
\end{equation}
They yield together with eqs.\ (2a)-(2c)
\begin{equation}
\frac{v^K}{v^p} =\frac{\bar{u}^{K^{+}}}{\bar{u}^p}=
\frac{g^K}{g^p}
\end{equation}
just as for the corresponding pion-proton relations in eq.\ (6).
Thus our basic predictions for the gluon and sea content of kaons
at the input scale $Q^2=\mu^2$ in Mellin $n$-moment space are
\begin{equation}
g^{K}=\frac{v^{K}}{v^p}\, g^p\;\;,\;\;
\bar{q}^{K}=\frac{v^{K}}{v^p}\, \bar{q}^p
\end{equation}
which is analogous to eq.\ (7). Taking again the input parton densities
of the proton from \cite{ref5}, only the total valence density of the
kaon, $v^K\equiv u_v^{K^{+}}+\bar{s}_v^{K^{+}}$, remains to be fixed.
In contrast to the pion, the constituent quarks have now different 
masses, i.e., $M_s>M_u$ so that the valence distribution 
$\bar{s}_v^{K^{+}}$ is expected to {\em{differ}} from
$u_v^{K^{+}}$ in being somewhat harder: $\bar{s}_v^{K^{+}}>u_v^{K^{+}}$
as $x\rightarrow 1$, i.e., the heavier $\bar{s}$ in $K^+$ should carry
more momentum than the lighter $u$ $(d)$. Unfortunately, the details
of this difference are not yet explored experimentally nor reliably
predicted theoretically [13-15]. The only experimental information 
available concerns $u_v^{K^{+}}$ which derives from the Drell-Yan 
process $K^- p \rightarrow \mu^+ \mu^- X$ at 
$4.1\,\mathrm{GeV}\le M_{\mu^+\mu^-}\le 8.5\,\mathrm{GeV}$ \cite{ref14}.
It indicates that $u_v^{K^{+}}<u_v^{\pi^{+}}$ for $x>0.6$ or
$u_v^{K^{+}}/u_v^{\pi^{+}}\rightarrow 1/2$ as $x\rightarrow 1$ at
$\langle Q^2\rangle = 20-40\,\mathrm{GeV}^2$. This requirement can be
easily accounted for by the ansatz 
$u_v^{K^{+}}(x,\mu^2)=N_u (1-x)^{\kappa} v^{\pi}(x,\mu^2)$ with
$\kappa$ being fitted to the (scarce) NA3 data \cite{ref14} and
$N_u$ follows from
\begin{equation}
\int_0^1 u_v^{K^{+}}(x,Q^2)\, dx = 
\int_0^1 \bar{s}_v^{K^{+}}(x,Q^2)\, dx =1\;\;\;.
\end{equation}
In analogy to eq.\ (9), we have furthermore
\begin{equation}
\int_0^1 x v^K(x,Q^2)\, dx = 
\int_0^1 x v^p(x,Q^2)\, dx =
\int_0^1 x v^{\pi}(x,Q^2)\, dx\;\;\;.
\end{equation}
In view of the absence of any experimental information about
$\bar{s}^{K^{+}}(x,Q^2)$, we take as our input
\begin{equation}
\bar{s}_v^{K^{+}}(x,\mu^2) = v^{\pi}(x,\mu^2)-u_v^{K^{+}}(x,\mu^2)
\end{equation}
which satisfies trivially the expectation discussed above as well as
the sum rules (17) and (18). Fitting now our ansatz for 
$u_v^{K^{+}}$ to the NA3 data \cite{ref14} yields
\begin{eqnarray}
u_{v,LO}^{K^{+}}(x,\mu_{LO}^2) &=& 0.541\, (1-x)^{0.17} 
v_{LO}^{\pi}(x,\mu_{LO}^2)\\
u_{v,NLO}^{K^{+}}(x,\mu_{NLO}^2) &=& 0.540\, (1-x)^{0.17}
v_{NLO}^{\pi}(x,\mu_{NLO}^2)\;\;\;.
\end{eqnarray}
The total momentum fractions carried by these LO and NLO light 
valence input densities are
\begin{equation}
\int_0^1 x u_{v,LO}^{K^{+}}(x,\mu_{LO}^2)\, dx =0.276\;\;,\;\;
\int_0^1 x u_{v,NLO}^{K^{+}}(x,\mu_{NLO}^2)\, dx = 0.267\;\;.
\end{equation}
Therefore the total momentum fractions carried by the heavy
strange input densities are, according to eqs.\ (19) and (12),
\begin{equation}
\int_0^1 x \bar{s}_{v,LO}^{K^{+}}(x,\mu_{LO}^2)\, dx =0.327\;\;,\;\;
\int_0^1 x \bar{s}_{v,NLO}^{K^{+}}(x,\mu_{NLO}^2)\, dx = 0.315\;\;.
\end{equation}

These LO input valence densities as well as the ones evolved to 
$Q^2=20\,\mathrm{GeV}^2$ are shown in fig.\ 5, and 
$u_v^{K^{+}}/u_v^{\pi^{+}}$ at $Q^2=20\,\mathrm{GeV}^2$ is 
compared with the NA3 data \cite{ref14} in fig.\ 6. 
The NLO valence densities are very similar to the LO ones shown
in fig.\ 5. Our expectations are similar to the ones derived from 
non-relativistic bound state (potential) and Nambu-Jona-Lasino-type 
models [13-15].
Due to our ansatz (19), i.e., $v^{K^{+}}\equiv u_v^{K^{+}}+
\bar{s}_v^{K^{+}}=v^{\pi^{+}}$, our predictions (16) for
$g^K(x,Q^2)$ and $\bar{q}^K(x,Q^2)$ become identical to the ones for
the pion in eq.\ (7), i.e., $g^K=g^{\pi}$ and $\bar{q}^K=\bar{q}^{\pi}$,
which are shown in figs.\ 1-3. It is thus obvious that also
$s^{K^{+}}=s^{\pi}$.

%%%%%%%%%%%%%%%%%%%
%  SECTION 4      %
%%%%%%%%%%%%%%%%%%%
%
\section{Discussion}
The present approach to the constituent quark model replaces the 
previous \cite{ref4} reliance on theoretically inferred constituent
wave functions \cite{ref1,ref4}. by experimentally extracted mesonic
valence quark distributions. It is shown that these distributions
together with the rather well known nucleon parton distributions
provide all the required information for predicting the gluon and sea
distributions within a meson. A confrontation with available 
experimental data supports the basic correctness of the underlying
constituent quark model and of our approach to its practical,
wave-function independent, implementation. Future improvements in
our knowledge of the nucleon and meson parton distributions are
required to test how reliable and correct the constituent quark model
actually is. 

A {\sc{Fortran}} package containing our LO and 
NLO$(\overline{\mathrm{MS}})$ pion densities can be obtained by
electronic mail on request.

%%%%%%%%%%%%%%%%%
%Acknowledgements
%%%%%%%%%%%%%%%%%
%
\section*{Acknowledgements}
The work has been supported in part by the
'Bundesministerium f\"{u}r Bildung, Wissenschaft, Forschung und
Technologie', Bonn.
%
%%%%%%%%%%%%%%%%%%%%%%%%%%%%
%

%
\newpage
%
%%%%%%%%%%%%%%%%%%
%  CAPTIONS
%%%%%%%%%%%%%%%%%%
\section*{Figure Captions}
\begin{description}
\item[Fig.1] The valence-like input distributions 
$xf^{\pi}(x,Q^2=\mu^2)$ with $f=v,$ $\bar{q}$, $g$ as compared to
those of ref.\ \cite{ref6} denoted by GRV$_{\pi}$. Note that
GRV$_{\pi}$ employs a vanishing SU(3)$_{flavor}$ symmetric 
$\bar{q}^{\pi}$ input. Similarly our present SU(3)$_{flavor}$ broken
sea densities refer to a vanishing $s^{\pi}$ input in eq.\ (5).
\item[Fig.2] The radiatively generated pionic gluon and sea-quark
distributions at various fixed values of $Q^2$ as compared to those of
ref.\ \cite{ref6} denoted by GRV$_{\pi}$. The predictions for the strange
sea density $s^{\pi}=\bar{s}^{\pi}$ are similar to the LO and NLO
GRV$_{\pi}$ results for $\bar{q}^{\pi}$.
\item[Fig.3] A detailed comparison of NLO pionic parton distributions
at $Q^2=20\,{\rm{GeV}}^2$. The stars (SMRS) refer to the distributions
of ref.\ \cite{ref9}. It should be noted that our present 
$\bar{q}^{\pi}\equiv\bar{u}^{\pi^{+}}=d^{\pi^{+}}$ (solid line), while
the GRV$_{\pi}$ and SMRS $\bar{q}^{\pi}\equiv\bar{u}^{\pi^{+}}=d^{\pi^{+}}=
s^{\pi}=\bar{s}^{\pi}$.
\item[Fig.4] A comparison of a sample of experimental Drell-Yan data
($\pi^-W$ reactions) \cite{ref10,ref9} with LO and NLO predictions based
on the present pion distributions together with the nucleon distributions
of \cite{ref5}.
\item[Fig.5] The LO valence distributions of the $K^+$ meson
at the input scale $Q^2=\mu^2$ and at $Q^2=20\,{\rm{GeV}}^2$ as
compared to the corresponding valence distribution of the pion.
\item[Fig.6] The ratio of the up-quark valence distributions of the
$K^+$ and $\pi^+$ mesons as compared with the corresponding 
experimental NA3 Drell-Yan data \cite{ref14}.
\end{description}
%
%%%%FIGURES
%
\newpage
\pagestyle{empty}
\vspace*{-0.6cm}
\hspace*{-1.1cm}
\epsfig{file=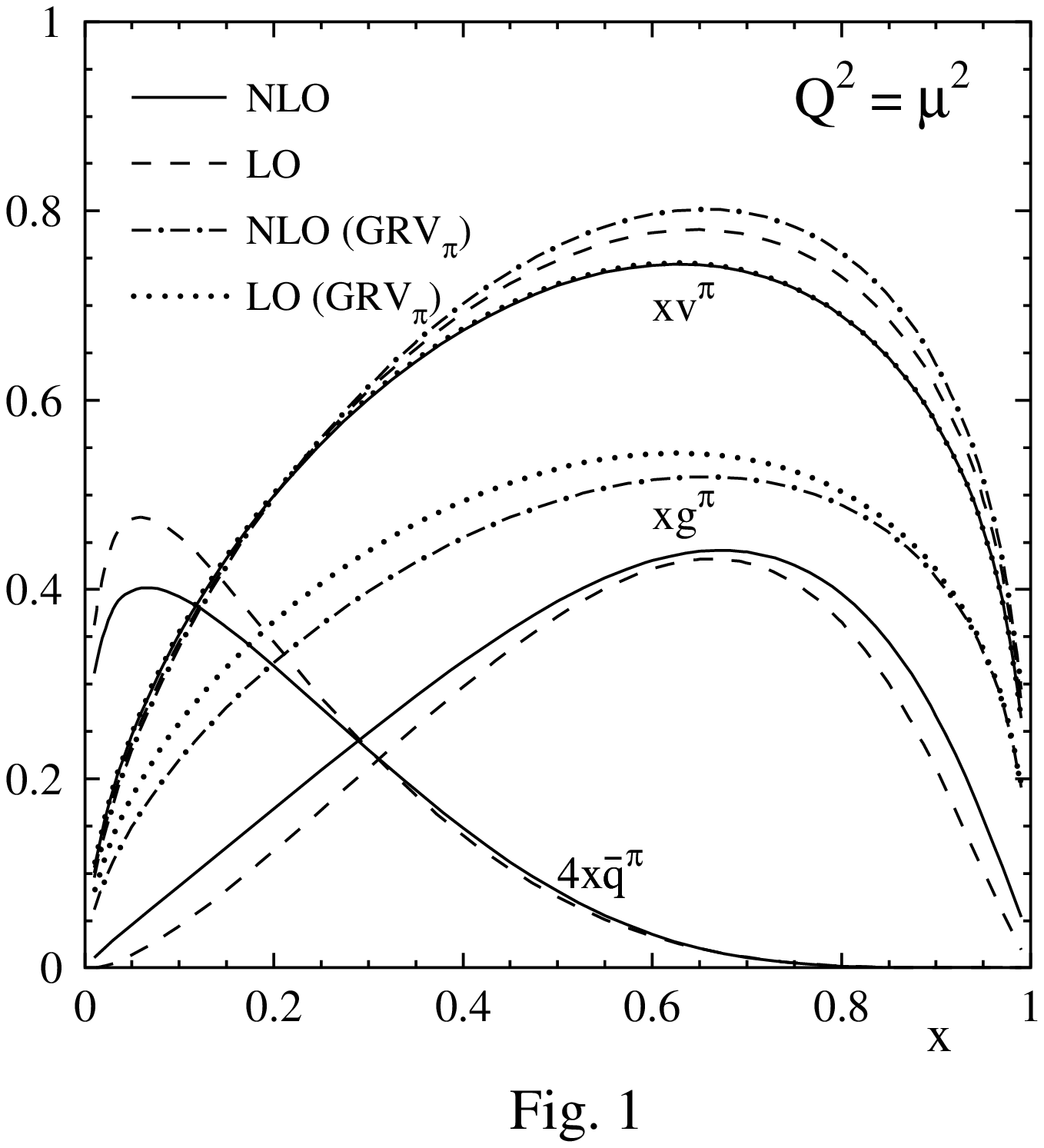}

\newpage
\vspace*{-1.3cm}
\hspace*{-0.9cm}
\epsfig{file=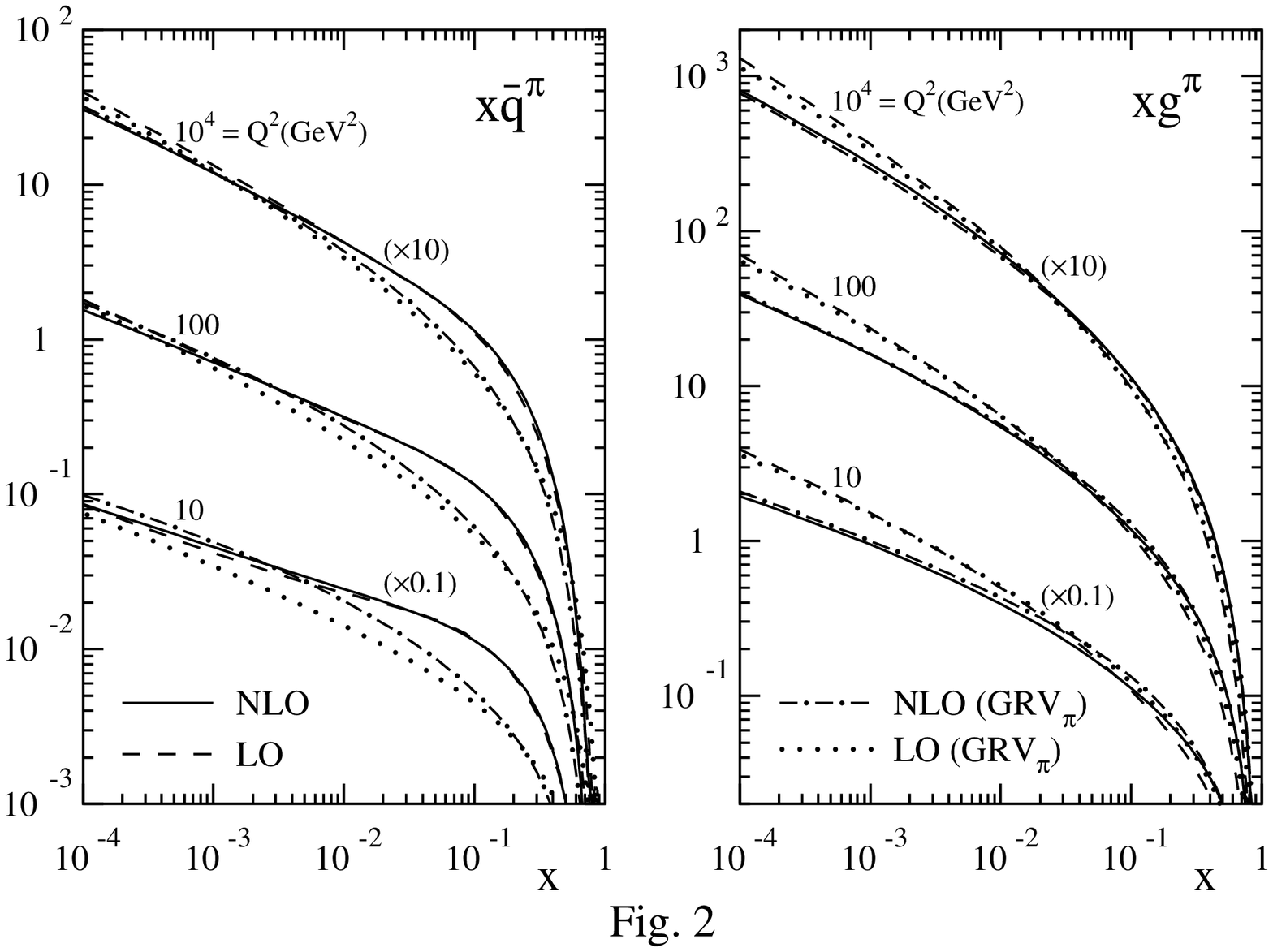,angle=90}

\newpage
\vspace*{-1.7cm}
\hspace*{-0.8cm}
\epsfig{file=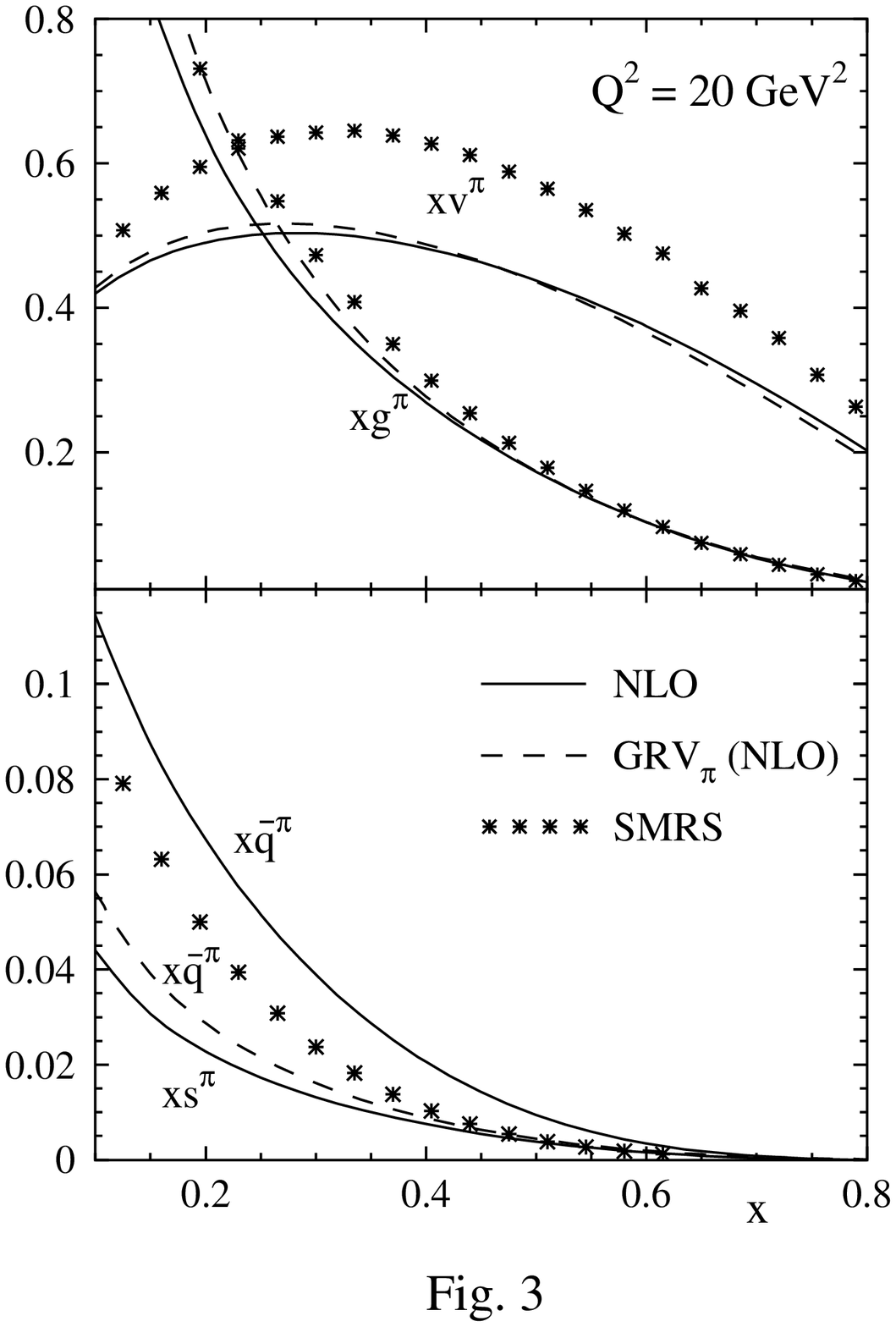}

\newpage
\vspace*{-2.0cm}
\hspace*{-0.8cm}
\epsfig{file=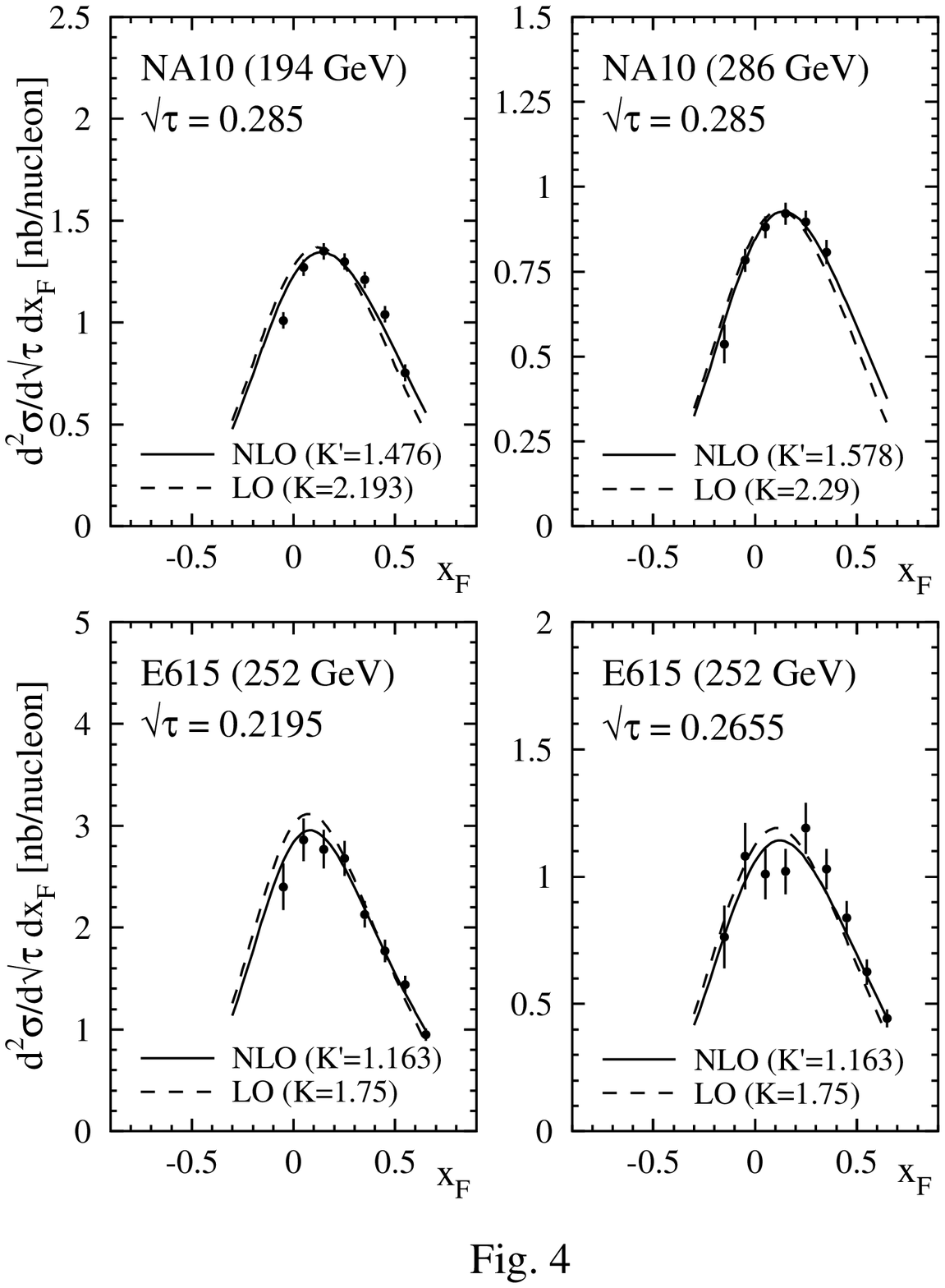}

\newpage
\vspace*{-0.6cm}
\hspace*{-1.1cm}
\epsfig{file=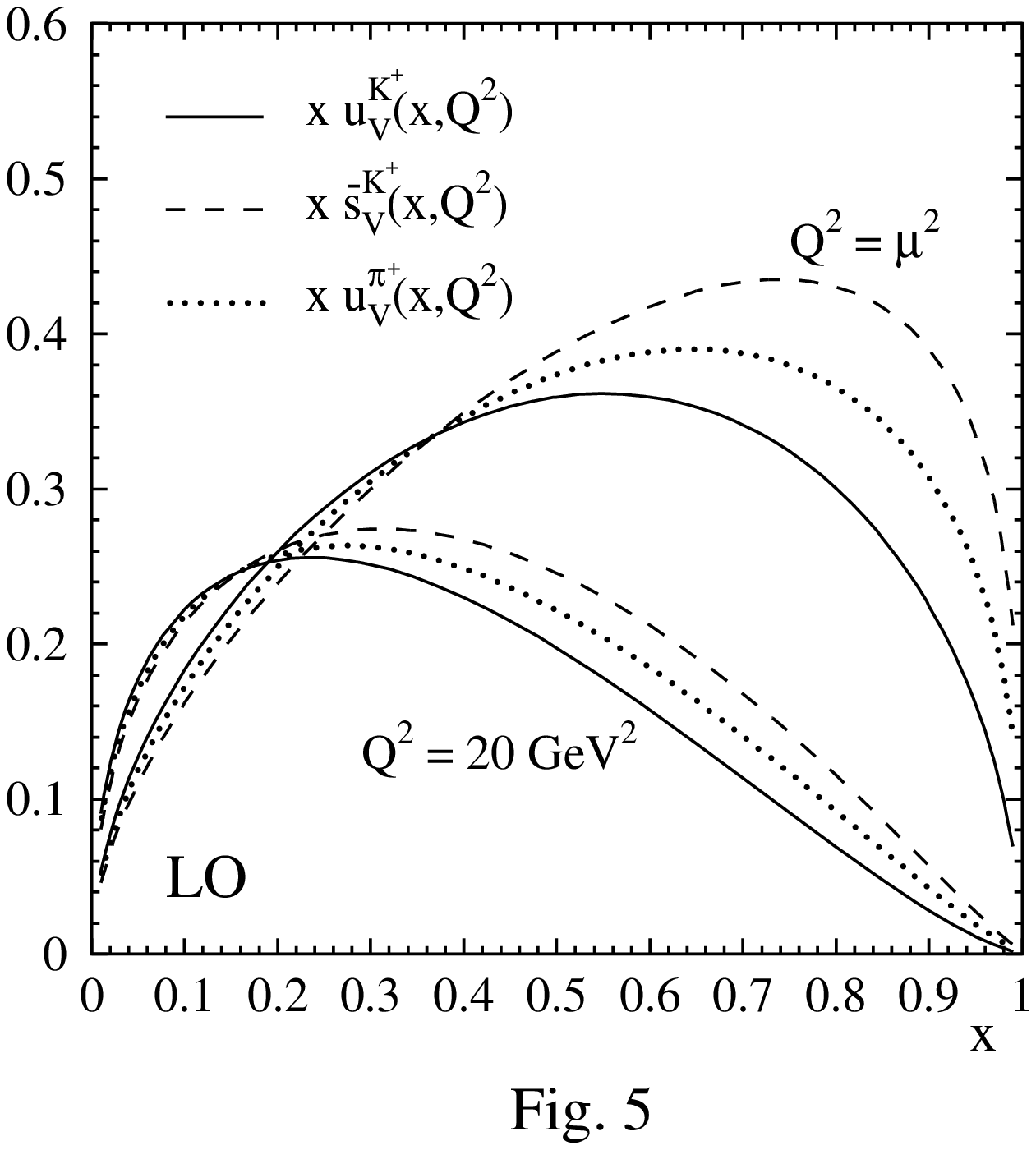}

\newpage
\vspace*{-0.6cm}
\hspace*{-1.1cm}
\epsfig{file=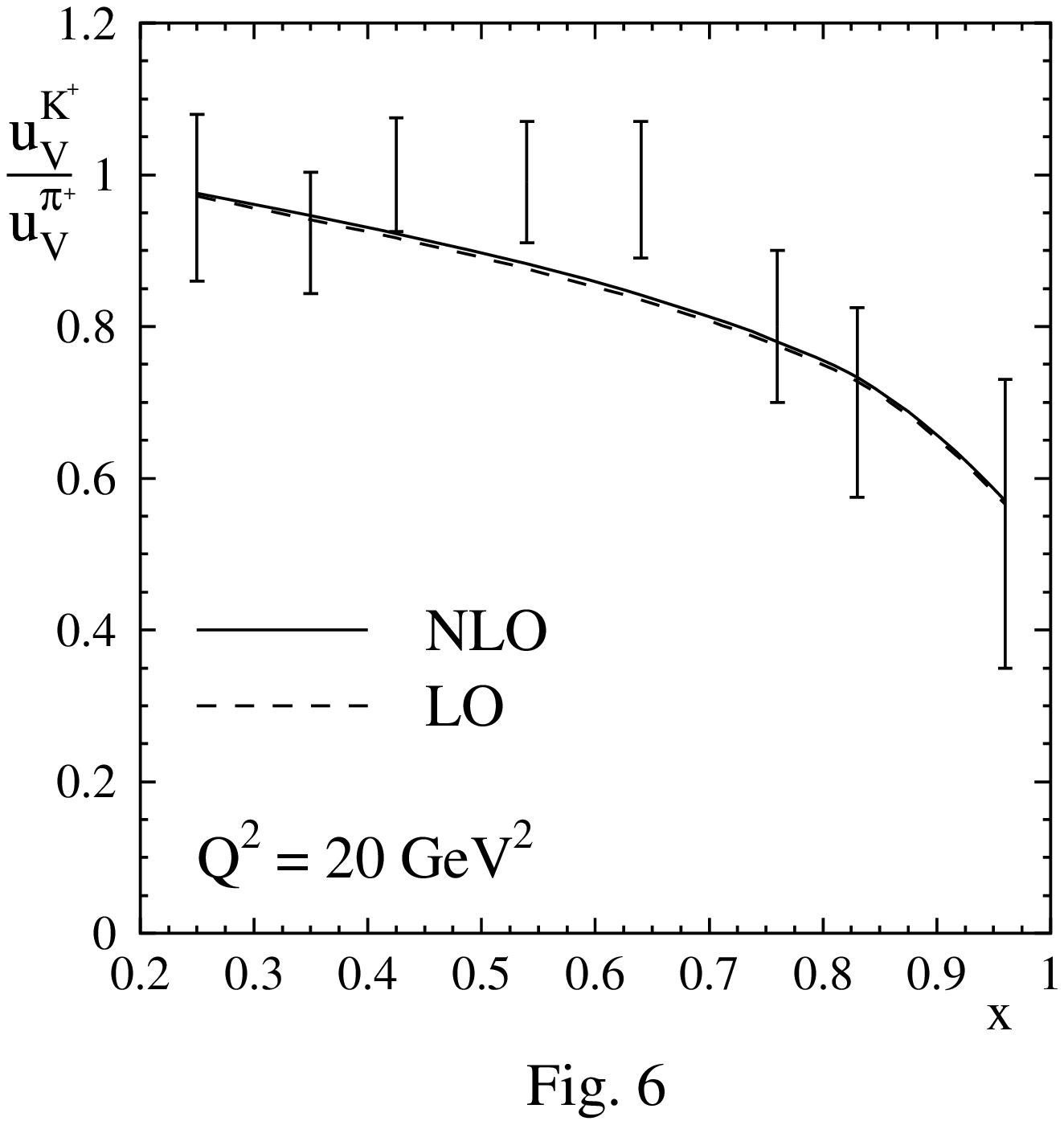}

\end{document}